\documentclass[a4paper,11pt]{article}
\usepackage{jheppub} 

\usepackage{graphicx}
\makeatletter
\g@addto@macro\@floatboxreset\centering
\makeatother

\arxivnumber{ } 

\makeatother

\begin{document}

\title{Multiparameter quantum estimation with a uniformly accelerated
Unruh-DeWitt detector}
\author{Shoukang Chang$^{1}$}
\author{Yashu Yang$^{1}$}
\author{Wei Ye$^{2}$}
\author{Yawen Tang$^{1}$}
\author{Hui Cao$^{1}$}
\author{Huan Zhang$^{3}$}
\author{Zunlue Zhu$^{1}$}
\author{Shaoming Fei$^{1,4}$}
\author{Xingdong Zhao$^{1}$}
\affiliation{$^{{\small 1}}$\textit{School of Physics, Henan Normal University, Xinxiang
453007, China}\\
$^{{\small 2}}$\textit{School of Information Engineering, Nanchang Hangkong
University, Nanchang 330063, China}\\
$^{{\small 3}}$\textit{Department of Physics, Weinan Normal University,
Wei'nan 714099, China}\\
$^{{\small 4}}$\textit{School of Mathematical Sciences, Capital Normal
University, Beijing 100048, China}}

\emailAdd{feishm@cnu.edu.cn}
\emailAdd{phyzhxd@gmail.com}

\abstract
{The uniformly accelerated Unruh-DeWitt detector serves as a fundamental
model in relativistic quantum metrology. While previous studies have mainly
concentrated on single-parameter estimation via quantum Cram\'{e}r--Rao
bound, the multi-parameter case remains significantly underexplored. In this
paper, we investigate the multiparameter estimation for a uniformly
accelerated Unruh-DeWitt detector coupled to a vacuum scalar field in both
bounded and unbounded Minkowski vacuum. Our analysis reveals that quantum
Cram\'{e}r--Rao bound fails to provide a tight error bound for the
two-parameter estimation involving the initial phase and weight parameters.
For this reason, we numerically compute two tighter error bounds, Holevo Cram%
\'{e}r--Rao bound and Nagaoka bound, based on a semidefinite program.
Notably, our results demonstrate that Nagaoka bound yields the tightest
error bound among all the considered error bounds, consistent with the
general hierarchy of multiparameter quantum estimation. In the case with a
boundary, we observe the introduction of boundary systematically reduces the
values of both Holevo Cram\'{e}r--Rao bound and Nagaoka bound, indicating an
improvement on the attainable estimation precision. These results offer
valuable insights on and practical guidance for advancing multiparameter
estimation in relativistic context.}

\keywords{: Unruh-DeWitt Detector, Multiparameter Quantum Estimation,
Minkowski Vacuum}

\maketitle

\section{Introduction}

Quantum parameter estimation (QPE) is a rapidly developing interdisciplinary
domain that bridges classical parameter estimation theory with quantum
mechanics \cite{1,2,3}. The fundamental objective of the QPE theory lies in
achieving the enhanced measurement precision for unknown parameters beyond
the capabilities of classical approaches, accomplished through the strategic
design of the QPE protocols using quantum entanglement resources \cite{4,5,6}%
, nonclassical states \cite{7,8} and quantum correlations \cite{9,10,11}. To
effectively evaluate the performance of the QPE protocols, the quantum Cram%
\'{e}r--Rao bound (CRB) is often used as a fundamental theoretical framework
renowned for determining the asymptotically achievable lower bounds on the
estimation precision \cite{12,13,14}. In this case, the inverse of the
quantum CRB, known as the quantum Fisher information, thus serves as a
fundamental metric quantifying the quantum state's sensitivity to minor
parameter variations \cite{15,16,17}. The quantum Fisher information has
transcended its initial application in QPE, emerging as an versatile and
analytical tool with widespread applications in quantum lidar \cite{18,19},
quantum telescopy \cite{20,21} and quantum thermometry \cite{22,23,24}.

Recently, the significant progress has been achieved in the applications of
QPE under the relativistic cases \cite{25,26,27,28,29}, including
acceleration \cite{28}, temperature \cite{30,31} and the Unruh-Hawking
effect \cite{32}. Among the pivotal applications of QPE in relativistic
contexts, the characterization of uniformly accelerating observers is
particularly significant, offering fundamental insights into the quantum
information processing in relativistic frameworks \cite{28,33,34,35}. For
instance, Zhao \textit{et al.} explored the quantum estimation of both
acceleration and temperature for a uniformly accelerated Unruh-DeWitt
detector coupled to a massless scalar field in the Minkowski vacuum \cite{28}%
. Their findings indicated that the optimal precision for acceleration
estimation is attained at specific acceleration values, also demonstrating
that the introduction of a boundary enhances the estimation precision of
both acceleration and temperature. Subsequently, Liu \textit{et al.}
investigated the estimation of the initial weight parameter, phase parameter
and inverse of acceleration for a uniformly accelerated Unruh-DeWitt
detector coupled to massless scalar field \cite{33}. Nevertheless, these
research contributions primarily employ the quantum CRB to address the
single-parameter estimation problem in a uniformly accelerated Unruh-DeWitt
detector, leaving the more complex realm of multiparameter estimation
largely unexplored. Consequently, the investigation of multiparameter
estimation in a uniformly accelerated Unruh-DeWitt detector remains an open
problem.

Generalizing from single-parameter to multiparameter quantum estimation
presents a nontrivial challenge \cite{36,37,38}. Unlike single-parameter
estimation, the optimal measurements for different parameters are often
incompatible, rendering the quantum CRB generally non-tight in
multiparameter scenarios \cite{36,37,38,39}. Theoretically, the quantum CRB
arises from quantizing the classical CRB using the symmetric logarithmic
derivative (SLD) \cite{40}, name as SLD-CRB. Nevertheless, the quantization
process of the classical CRB is not unique \cite{41}. Utilizing the right
logarithmic derivative (RLD) yields the RLD-CRB \cite{42}, which also
suffers from the potential non-tightness as its optimal estimators may not
correspond to the physical realizable positive-operator-valued measures
(POVMs) \cite{43,44}. In order to tackle this problem, researchers often
resort to the Holevo Cram\'{e}r--Rao bound (HCRB), which can provide a
tighter precision limit than the SLD-CRB and RLD-CRB \cite{36,37,45}.

Generally, the HCRB can be not only available through executing collective
measurements on infinitely many copies of the quantum state in the
asymptotic case \cite{37,45}, but also achieved by the single-copy
measurements for the pure state \cite{46} and displacement estimation with
Gaussian states \cite{47,48,49,50,51}. Despite its fundamental significance,
the application of the HCRB in multiparameter estimation is hindered by the
computational intractability, which involves a complex optimization problem
over a set of observables. Recently, this optimization problem has been
formulated as a semidefinite program (SDP) for the finite-dimensional\textbf{%
\ }\cite{52} and infinite-dimensional Gaussian systems \cite{53}, rendering
the numerical evaluation relatively straightforward. Crucially, the HCRB's
asymptotic achievability requirement for collective measurements \cite%
{36,37,45} poses significant experimental challenges \cite{54}, highlighting
the need for tighter bounds under separable, single-copy measurements. For
two-parameter qubit estimation, the tight Nagaoka bound (NB) fulfills this
need \cite{55}, but its extension to more parameters, the Nagaoka-Hayashi
bound (NHB), is not generally tight \cite{54,56,57,58}. Similar to the HCRB,
the computation of both the NB and the NHB involves a non-trivial
optimization problem, which can be numerically solved using the SDP \cite%
{54,59,60}.

In this paper, we investigate the multiparameter estimation for a uniformly
accelerated two-level atom system, known as an Unruh-DeWitt detector,
interacting with a vacuum scalar field in both bounded and unbounded
Minkowski vacuum. For the unbounded case, we focus on the joint estimation
of atom's initial phase and weight parameters. Our results reveal that the
corresponding SLD operators are noncommuting and the Uhlmann curvature
matrix \cite{36,61} is non-zero, implying that the SLD-CRB can not provide
an asymptotically tight error bound. Similarly, the RLD-CRB is also
generally non-tight. Consequently, we numerically compute the HCRB and NB
using the SDP. Our results verify that the NB consistently provides the
tightest achievable precision bound among the SLD-CRB, RLD-CRB, HCRB and NB,
which aligns with the general hierarchy of multiparameter quantum
estimation. Notably, while these error bounds vary monotonically with the
inverse of acceleration and proper time, they exhibit non-monotonic behavior
with respect to the weight parameter. Crucially, we observe a significant
competition and crossover in tightness between the SLD-CRB and RLD-CRB,
particularly as the proper time and the inverse of acceleration vary. By
extending the results to the case of three parameters (phase, weight and the
inverse of acceleration) in the two-level atom system, our numerical
calculations show that the NHB consistently produces the largest values,
confirming its status as the tightest bound. The RLD-CRB and HCRB are
numerically identical and both exceed the SLD-CRB, indicating that they
offer asymptotically tight precision limits, whereas the SLD-CRB shows the
weakest tightness. In the case with a boundary, we have also examined both
two-parameter and three-parameter estimation in the two-level atom system.
While observing similar trends to the unbounded case, we find that the
introduction of the boundary reduces the numerical values of HCRB, NB and
NHB, signifying a notable enhancement in the attainable estimation precision.

The remainder of this paper is arranged as follows. In Sec. II, we review
some results of multiparameter quantum estimation theory. In Sec. III, we
explore the multiparameter estimation problem for a uniformly accelerated
two-level atom system. Finally, our main conclusions are drawn in the last
section.

\section{Preliminaries}

In this section, we recall some basic elements in theory of multiparameter
quantum estimation. Consider a generic quantum statistical model $\hat{\rho}%
_{\theta }$ parameterized by multiple unknown parameters $\theta=(\theta
_{1},...,\theta _{d})^{\text{T}}$ to be estimated, where $T$ denotes the
transpose. In order to extract the physical information regarding these
unknown parameters, we implement the POVMs on the quantum statistical model $%
\hat{\rho}_{\theta }.$ The corresponding conditional probability associated
with measurement outcome $k$ is governed by the Born's rule \cite{62}, 
\begin{equation}
P(k|\theta )=\text{Tr(}\hat{\rho}_{\theta }\hat{\Pi}_{k}\text{),}  \label{1}
\end{equation}%
where Tr($\cdot $) denotes the trace of an operator in Hilbert space, $\hat{%
\Pi}_{k}$ is the $k$th measurement operator of the POVM, satisfying $\hat{\Pi%
}_{k}\geq 0$ and $\sum_{k}\hat{\Pi}_{k}$=$I$, with $I$ being the identity
operator.

The estimator function $\check{\theta}(k)$ serves as a tool for deducing the
values of the unknown parameters based on the measurement outcomes. The
effectiveness of the estimator function $\check{\theta}(k)$ in parameter
estimation can be characterized by the mean square error matrix, 
\begin{equation}
\Sigma _{\theta }(\hat{\Pi}_{k},\check{\theta}(k))\text{=}%
\sum\limits_{k}P(k|\theta )(\check{\theta}(k)-\theta )(\check{\theta}%
(k)-\theta )^{\text{T}}.  \label{2}
\end{equation}%
Within the frequentist multiparameter estimation framework, the following
locally unbiasedness constraint\textbf{\ }condition, 
\begin{eqnarray}
\sum\limits_{k}(\check{\theta}_{\mu }(k)-\theta _{\mu })P(k|\theta ) &=&0, 
\notag \\
\sum\limits_{k}\check{\theta}_{\mu }(k)(\left. \partial P(k|\theta )\right/
\partial \theta _{v}) &=&\delta _{\mu v},  \label{3}
\end{eqnarray}%
is conventionally imposed on the estimator function $\check{\theta}(k)$ to
address the minimization problem associated with the trace of the mean
square error matrix. Under these conditions, a lower bound for the mean
square error matrix, i.e., the matrix CRB \cite{40,63}, is given by%
\begin{equation}
\Sigma _{\theta }(\hat{\Pi}_{k},\check{\theta}(k))\geq F^{-1},  \label{4}
\end{equation}%
where $F$ is the classical Fisher information matrix.

The matrix CRB fundamentally characterizes the minimum achievable mean
squared error matrix for a given measurement scheme under the optimal
classical data processing. This theoretical limit can be asymptotically
attained by using an appropriate and efficient estimator. To attain the
ultimate precision limits in multiparameter estimation, the matrix CRB has
been quantized, giving rise to two distinct quantum versions. A renowned
quantum lower bound for the mean square error matrix is related to the real
symmetric quantum Fisher information matrix, with elements \cite{64,65} 
\begin{equation}
J_{uv}^{S}=\frac{1}{2}\text{Tr}\left[ \hat{\rho}_{\theta }(\hat{L}_{u}^{S}%
\hat{L}_{v}^{S}+\hat{L}_{v}^{S}\hat{L}_{u}^{S})\right] ,  \label{5}
\end{equation}%
where the SLD operators $\hat{L}_{u}^{S}$ satisfy the Lyapunov equation $%
\partial \hat{\rho}_{\theta }/\partial \theta _{u}$=$\left. (\hat{L}_{u}^{S}%
\hat{\rho}_{\theta }+\hat{\rho}_{\theta }\hat{L}_{u}^{S})\right/ 2$. Another
important one is relevant to the RLD quantum Fisher information matrix,
whose elements are \cite{42} 
\begin{equation}
J_{uv}^{R}=\text{Tr}\left[ \left( \hat{L}_{u}^{R}\right) ^{\dagger }\hat{\rho%
}_{\theta }\hat{L}_{v}^{R}\right] ,  \label{6}
\end{equation}%
with the RLD operators $\hat{L}_{u}^{R}$ defined by $\partial \hat{\rho}%
_{\theta }/\partial \theta _{u}$=$\hat{\rho}_{\theta }\hat{L}_{u}^{R}.$ To
quantify the tightness of these error bounds, one derives the following
scalar forms for the SLD-CRB and the RLD-CRB \cite{36,37,45}, 
\begin{eqnarray}
C_{\theta }^{S} &=&\text{tr}[(J^{S})^{-1}],  \notag \\
C_{\theta }^{R} &=&\text{tr}[\text{Re}(J^{R})^{-1}]+\left\Vert \text{Im}%
(J^{R})^{-1}\right\Vert _{1},  \label{7}
\end{eqnarray}%
where tr[$\cdot $] represents the trace of finite dimensional $d\times d$
matrices, $\left\Vert A\right\Vert _{1}$=tr($\sqrt{A^{\dagger }A}$) is the
trace norm and Re$\left( \cdot \right) $ denotes the real part of a matrix.
Unlike the single-parameter estimation, the SLD-CRB is generally not tight
due to the incompatibility of the optimal measurements for different
parameters. Likewise, the RLD-CRB is also generally not tight, since the
optimal estimators for the RLD-CRB may not be physical POVMs.

Holevo proposed a tighter scalar bound known as the HCRB, which is defined
via the following minimization problem \cite{37,66}, 
\begin{equation}
C_{\theta }^{H}=\min_{\hat{X}}\left[ \text{tr[Re}Z[\hat{X}]\text{]}%
+\left\Vert \text{Im}Z[\hat{X}]\right\Vert _{1}\right] ,  \label{8}
\end{equation}%
where $\hat{X}$=($\hat{X}_{1},...\hat{X}_{d}$)$^{\text{T}}$ is a vector of
Hermitian operators satisfying the locally unbiased conditions, 
\begin{eqnarray}
\text{Tr}\left[ \hat{\rho}_{\theta }\hat{X}_{u}\right] &=&0,  \notag \\
\text{Tr}\left[ \hat{X}_{u}\partial \hat{\rho}_{\theta }/\partial \theta _{u}%
\right] &=&\delta _{uv},  \label{9}
\end{eqnarray}%
$Z[\hat{X}]$ is a $d\times d$ Hermitian matrix with entries $Z[\hat{X}]_{uv}=%
\text{Tr}\left[ \hat{\rho}_{\theta }\hat{X}_{u}\hat{X}_{v}\right]$.

In fact, the HCRB\textbf{\ }is tighter than both the SLD-CRB and the
RLD-CRB, and can be thus achieved by performing a collective measurement
over infinitely many copies of quantum states \cite{36,37,45}. Recently, it
has been demonstrated that the HCRB can be considered as the upper bound of
SLD-CRB \cite{36,37} 
\begin{equation}
C_{\theta }^{S}\leq C_{\theta }^{H}\leq C_{\theta }^{U}\leq 2C_{\theta }^{S},
\label{10}
\end{equation}%
where we have defined the upper bound $C_{\theta }^{U}$=$C_{\theta
}^{S}+\left\Vert (J^{S})^{-1}D(J^{S})^{-1}\right\Vert _{1}$, with $D$ being
the mean Uhlmann curvature matrix \cite{36,59} given by the entries 
\begin{equation}
D_{uv}=-\frac{i}{2}\text{Tr[}\hat{\rho}_{\theta }[\hat{L}_{u}^{S},\hat{L}%
_{v}^{S}]\text{].}  \label{11}
\end{equation}%
The SLD-CRB is tight and $C_{\theta }^{S}$=$C_{\theta }^{H}$ when $[\hat{L}%
_{u}^{S},\hat{L}_{v}^{S}]$=$0,$ $\forall u,v.$ Meanwhile, we can always
identify a set of common eigenstates corresponding to these commuting
symmetric logarithmic derivative operators, which can serve as the POVM
measurement basis to saturate the SLD-CRB through single-copy measurements.
Moreover, there are some special quantum states such that $[\hat{L}_{u}^{S},%
\hat{L}_{v}^{S}]\neq 0$ and the mean Uhlmann curvature matrix $D$ is a zero
matrix. One can also derive that $C_{\theta }^{S}=C_{\theta }^{H}$ can be
saturated asymptotically by implementing the collective measurements.

However, the implementation of collective measurements remains
experimentally challenging with current technological capabilities \cite{54}%
. For this reason, Nagaoka introduced a more informative scalar bound (NB)
for two-parameter estimation \cite{55}, 
\begin{eqnarray}
C_{\theta }^{N} &=&\min_{\hat{X}}\{\text{Tr}[\hat{\rho}_{\theta }\hat{X}_{1}%
\hat{X}_{1}+\hat{\rho}_{\theta }\hat{X}_{2}\hat{X}_{2}]  \notag \\
&&+\text{TrAbs}[\hat{\rho}_{\theta }[\hat{X}_{1},\hat{X}_{2}]]\},  \label{12}
\end{eqnarray}
where TrAbs$[\hat{K}]$ denotes the sum of the absolute values of the
eigenvalues of the operator $\hat{K}.$ The NB was proven to be a tight
scalar bound for two-parameter estimation \cite{67}. In order to estimate
more than two parameters, we will invoke the NHB \cite{54}, which is
expressed as%
\begin{eqnarray}
C_{\theta }^{N} &=&\min_{\hat{L},\text{ }\hat{X}}\{\left. \mathbf{Tr[}\hat{S}%
_{\theta }\hat{L}\mathbf{]}\right\vert \hat{L}_{uv}\text{=}\hat{L}_{vu}\text{
Hermitian,}  \notag \\
&&\hat{L}\geq \hat{X}\hat{X}^{\text{T}}\},  \label{13}
\end{eqnarray}%
where $\hat{S}_{\theta }$=$1_{d}\otimes \hat{\rho}_{\theta }$ exists in an
expanded classical-quantum Hilbert space$,$ $1_{d}$ is\textbf{\ }the $%
d\times d$ identity matrix, $\hat{L}$ is the $d\times d$ matrix of Hermitian
operators, the symbol $\mathbf{Tr}$[$\cdot $] represents\textbf{\ }the trace
over both classical and quantum systems. For brevity of notation, we will
utilize the symbol $C_{\theta }^{N}$ to denote both the NB and the NHB. It
is noteworthy that Gill and Massar proposed an alternative bound \cite{68}.
Nevertheless, since this bound is generally less tight compared with the
NHB, we have chosen to exclude it from our discussion in this paper.
Theoretically, the most informative bound can always be defined as the
minimal scalar CRB optimized over all possible POVMs \cite{36,69,70,71},%
\begin{equation}
C_{\theta }^{MI}=\min_{\text{POVM}}[\text{tr}[F^{-1}]],  \label{14}
\end{equation}%
which satisfies the following chain of inequalities, 
\begin{eqnarray}
\text{tr}[\Sigma _{\theta }(\hat{\Pi}_{k},\check{\theta}(k))] &\geq
&C_{\theta }^{MI}\geq C_{\theta }^{N}\geq C_{\theta }^{H}  \notag \\
&\geq &\max [C_{\theta }^{S},C_{\theta }^{R}].  \label{15}
\end{eqnarray}%
It is worth emphasizing that the SLD-CRB, the HCRB, the NHB and the most
informative bound are numerically the same for the single-parameter
estimation \cite{70,71,72}. For the two-parameter estimation, the NB is a
tight scalar bound, thereby showing the equivalent numerical results with
the most informative bound \cite{70,71,72}. Furthermore, when estimating any
number of parameters using pure quantum states, the HCRB and the NHB are
numerically equal \cite{70,71,72}.

\section{Multiparameter quantum estimation in a two-level atom system}

In this section, we evaluate the ultimate bounds for multiparameter quantum
estimation by systematically analyzing a uniformly accelerated two-level
atom, i.e., an Unruh-DeWitt detector, interacting with a vacuum scalar field
in both bounded and unbounded Minkowski vacuum, and derive the five
fundamental precision limits, SLD-CRB, RLD-CRB, HCRB, NB and NHB.

In a two-level atom system, a quantum state $\hat{\rho}$ can typically be
expressed in Bloch representation \cite{62}, 
\begin{equation}
\hat{\rho}=\frac{1}{2}\left( I+\sum\limits_{j=1}^{3}\omega _{j}\hat{\sigma}%
_{j}\right) ,  \label{16}
\end{equation}%
where $(\omega _{1},\omega _{2},\omega _{3})$ denotes the Bloch vector and $%
\hat{\sigma}_{j}$ are the standard Pauli matrices. We consider such a
two-level atom system coupled to a fluctuating vacuum scalar field in the
Minkowski vacuum. This physical model posits that the behavior of the
two-level atom is the same as the one of an open system, i.e., a system
immersed in an external environment field, where the vacuum fluctuations of
the quantum field constitute the environmental degrees of freedom \cite%
{28,31}. The complete dynamics of this coupled system (atom plus vacuum
scalar field) is governed by the total Hamiltonian \cite{28,31} 
\begin{equation}
\hat{H}=\hat{H}_{s}+\hat{H}_{f}+\hat{H}_{I},  \label{17}
\end{equation}%
where $\hat{H}_{s}$=$\left. \text{%
h{\hskip-.2em}\llap{\protect\rule[1.1ex]{.325em}{.1ex}}{\hskip.2em}%
}\omega _{0}\hat{\sigma}_{3}\right/ 2$ is the Hamiltonian of the two-level
atom with $\omega _{0}$ denotes the energy level spacing of the atom, $\hat{H%
}_{f}$ is the Hamiltonian of the vacuum scalar field\textbf{, }and\textbf{\ }%
$\hat{H}_{I}=\mu (\hat{\sigma}_{+}+\hat{\sigma}_{-})\hat{\phi}(t,\mathbf{x}) 
$ is the interaction Hamiltonian between the two-level atom and the vacuum
scalar field, with $\mu $ being the coupling constant, $\hat{\sigma}_{+}$
and $\hat{\sigma}_{-}$ the atomic raising and lowering operators,
respectively, and $\hat{\phi}(t,\mathbf{x})$ the scalar field operator.

Assume that the initial total density matrix of the coupled system takes $%
\hat{\rho}_{\text{tot}}(0)$=$\hat{\rho}(0)\otimes \left\vert 0\right\rangle
\left\langle 0\right\vert ,$ where $\hat{\rho}(0)$ is the initial reduced
density matrix of the two-level atom and $\left\vert 0\right\rangle
\left\langle 0\right\vert $ is the vacuum state of the scalar field. If the
interaction between the two-level atom and the vacuum scalar field is weak,
the corresponding reduced density matrix $\hat{\rho}(\tau )$ obeys an
equation in the Kossakowski-Lindblad form \cite{73,74}, 
\begin{equation}
\frac{\partial \hat{\rho}(\tau )}{\partial \tau }=-\frac{i}{\text{%
h{\hskip-.2em}\llap{\protect\rule[1.1ex]{.325em}{.1ex}}{\hskip.2em}%
}}[\hat{H}_{\text{eff}},\hat{\rho}(\tau )]+\text{\L [}\hat{\rho}(\tau )\text{%
],}  \label{18}
\end{equation}%
where $\tau $ is the proper time, the effective Hamiltonian $\hat{H}_{\text{%
eff}}$, by absorbing the Lamb shift term, is given by \cite{28,31,35}, 
\begin{eqnarray}
\hat{H}_{\text{eff}} &=&\frac{1}{2}\text{%
h{\hskip-.2em}\llap{\protect\rule[1.1ex]{.325em}{.1ex}}{\hskip.2em}%
}\Omega \sigma _{3}  \notag \\
&=&\frac{\text{%
h{\hskip-.2em}\llap{\protect\rule[1.1ex]{.325em}{.1ex}}{\hskip.2em}%
}}{2}\left\{ \omega _{0}+\frac{i}{2}[K(-\omega _{0})-K(\omega _{0})]\right\}
\sigma _{3}  \label{19}
\end{eqnarray}%
with $\Omega $ being the renormalized energy gap, and the Lindblad term 
\begin{equation}
\text{\L [}\hat{\rho}\text{]}=\frac{1}{2}\sum\limits_{i,j=1}^{3}a_{ij}[2\hat{%
\sigma}_{j}\hat{\rho}\hat{\sigma}_{i}-\hat{\sigma}_{i}\hat{\sigma}_{j}\hat{%
\rho}-\hat{\rho}\hat{\sigma}_{i}\hat{\sigma}_{j}],  \label{20}
\end{equation}%
with the coefficients $a_{ij}$ of the Kossakowski matrix \cite{28,31,35}
given by $a_{ij}=A\delta _{ij}-iB\varepsilon _{ijk}\delta _{k3}-A\delta
_{i3}\delta _{j3}$ ($\delta _{ij}$\ and $\varepsilon _{ijk}$ respectively
represent the Kronecker Delta and the Levi-Civita symbol), 
\begin{eqnarray}
A &=&\frac{\mu ^{2}}{4}[G(\omega _{0})+G(-\omega _{0})],  \notag \\
B &=&\frac{\mu ^{2}}{4}[G(\omega _{0})-G(-\omega _{0})].  \label{21}
\end{eqnarray}
The $G(\omega _{0})$ and $K(\omega _{0})$ are given by 
\begin{eqnarray}
G(\omega _{0}) &=&\int_{-\infty }^{\infty }d\Delta \tau e^{i\omega
_{0}\Delta \tau }G^{+}(\Delta \tau ),  \notag \\
K(\omega _{0}) &=&\frac{P}{\pi i}\int_{-\infty }^{\infty }d\omega \frac{%
G(\omega )}{\omega -\omega _{0}},  \label{22}
\end{eqnarray}%
where $\Delta \tau $=$\tau -\tau ^{\prime }$, $P$ is the principle value,
and $G^{+}(\Delta \tau )$ is given by the two-point correlation function for
the scalar field, $G^{+}(x,x^{\prime })=\left\langle 0\right\vert \hat{\phi}%
(t,\mathbf{x})\hat{\phi}(t^{\prime },\mathbf{x}^{\prime })\left\vert
0\right\rangle $ \cite{28,35}.

If we choose the initial state of the two-level atom system as $\left\vert
\psi (0)\right\rangle $=$\cos $($\theta /2$)$\left\vert 1\right\rangle
+e^{i\phi }\sin $($\theta /2$)$\left\vert 0\right\rangle ,$ we can derive
the time-dependent reduced density matrix \cite{28,31,35} 
\begin{equation}
\hat{\rho}(\tau )=\frac{1}{2}\left( I+\sum\limits_{j=1}^{3}\omega _{j}(\tau
)\sigma _{j}\right) ,  \label{23}
\end{equation}%
where%
\begin{eqnarray}
\omega _{1}(\tau ) &=&\sin \theta \cos (\Omega \tau +\phi )e^{-2A\tau }, 
\notag \\
\omega _{2}(\tau ) &=&\sin \theta \sin (\Omega \tau +\phi )e^{-2A\tau }, 
\notag \\
\omega _{3}(\tau ) &=&\cos \theta e^{-4A\tau }-\frac{B}{A}(1-e^{-4A\tau }).
\label{24}
\end{eqnarray}

\subsection{Multiparameter quantum estimation without a boundary}

Let us begin with the multiparameter estimation for a uniformly accelerated
two-level atom system coupled to a vacuum scalar field in the Minkowski
vacuum \cite{28,75}. For the convenience of the following discussion and
analysis, we utilize natural units $c$=%
h{\hskip-.2em}\llap{\protect\rule[1.1ex]{.325em}{.1ex}}{\hskip.2em}%
=$k_{B}$=$1.$ The trajectory of this two-level atom system can be described
as follows \cite{28,75}, 
\begin{eqnarray}
t(\tau ) &=&\frac{1}{a}\sinh (a\tau ),  \notag \\
x(\tau ) &=&\frac{1}{a}\cosh (a\tau ),  \notag \\
y(\tau ) &=&y_{0},  \notag \\
z(\tau ) &=&z_{0},  \label{25}
\end{eqnarray}%
where $a$ is the acceleration of this two-level atom system. The
corresponding two-point correlation function for the vacuum scalar field in
the Minkowski vacuum is given by \cite{28,75}, 
\begin{eqnarray}
&&G^{+}(x,x^{\prime })_{0}  \notag \\
&=&-\frac{1/(4\pi ^{2})}{(t-t^{\prime }-i\epsilon )^{2}-(x-x^{\prime
})^{2}-(y-y^{\prime })^{2}-(z-z^{\prime })^{2}}  \notag \\
&=&-\frac{a^{2}}{16\pi ^{2}\sinh ^{2}\left( \frac{a\Delta \tau }{2}%
-i\epsilon \right) },  \label{26}
\end{eqnarray}%
where we have used Eq. (\ref{25}) in the last equality.

Based on Eq. (\ref{22}), we can derive the Fourier transformation of the
two-point correlation function \cite{28,75}, 
\begin{equation}
G(\omega _{0})_{0}=\frac{\omega _{0}}{2\pi (1-e^{-2\pi \omega _{0}/a})},
\label{27}
\end{equation}%
which allows us to determine the coefficients in the Kossakowski matrix
according to Eq. (\ref{21}), 
\begin{eqnarray}
A_{0} &=&\frac{\Gamma _{0}}{4}\coth \frac{\pi \omega _{0}}{a},  \notag \\
B_{0} &=&\frac{\Gamma _{0}}{4},  \label{28}
\end{eqnarray}%
where $\Gamma _{0}$=$\mu ^{2}\omega _{0}/2\pi $ denotes the spontaneous
emission rate. 

\begin{figure}[tbp]
\centering
\includegraphics[width=0.5\linewidth]{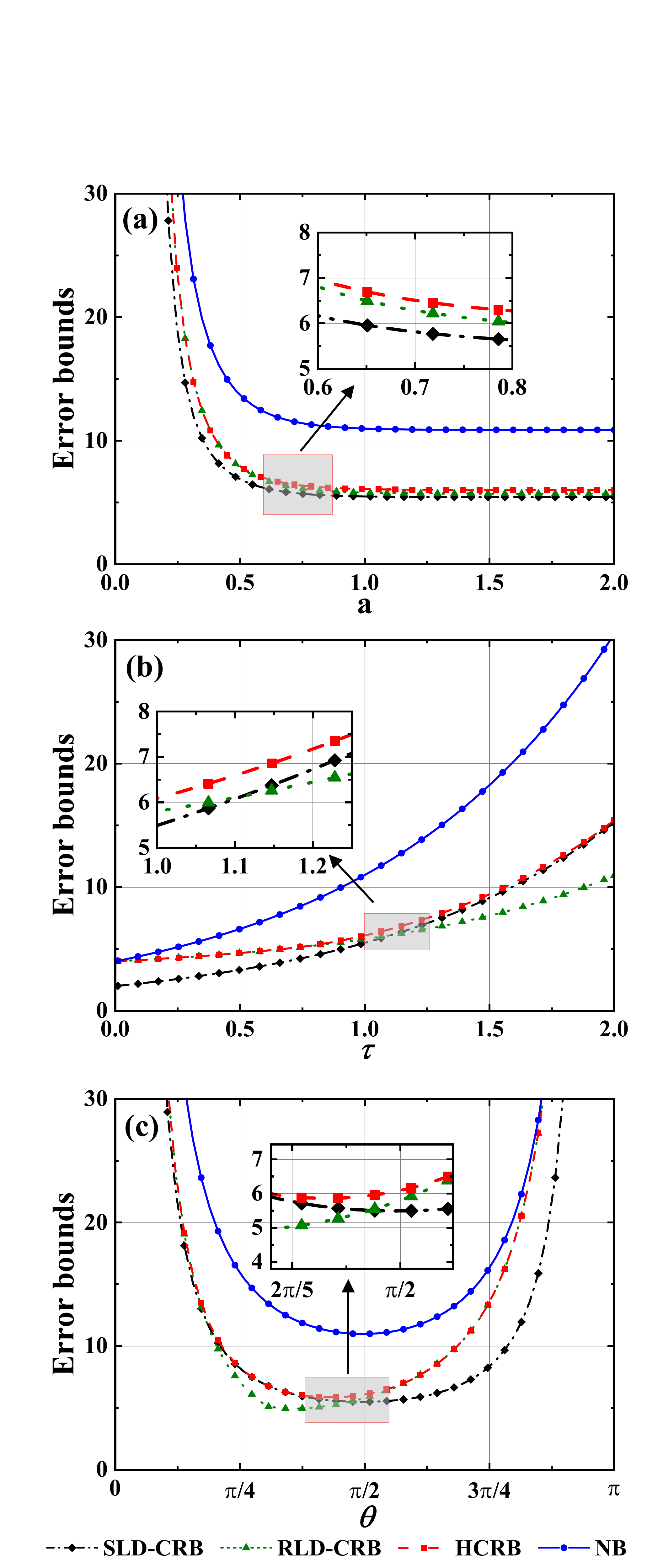}
\caption{Error bounds as a function of (a) the inverse of acceleration $a$ 
with $\theta = \pi/2$, $\tau = 0.4$, and $z = 0.5$; 
(b) the proper time $\tau$ with $\theta = \pi/2$, $a = 0.2$ and $z = 0.5$;
(c) the weight parameter $\theta$ with $\tau = 0.4$, $a = 0.2$ and $z = 0.5$.}
\label{fig:error_bounds}
\end{figure}

Therefore, substituting Eq. (\ref{28}) into Eq. (\ref{23}), one gets the
time-dependent reduced density matrix of the uniformly accelerated two-level
atom system. In the following discussion and analysis, we adopt the
transformations $\tau \rightarrow \tilde{\tau}$=$\Gamma _{0}\tau $ and $%
a\rightarrow $ $\tilde{a}$=$\omega _{0}/a.$ For convenience, $\tilde{\tau}$
and $\tilde{a}$ will be rewritten as $\tau $ and $a,$ respectively. We first
consider a two-parameter estimation involving the initial weight parameter $%
\theta $ and phase parameter $\phi $ of the two-level atom system$.$
Obviously, the symmetric logarithmic derivative operators $\hat{L}_{\theta
}^{S}$ and $\hat{L}_{\phi }^{S}$ are non-commutative. Furthermore, using Eq.
(\ref{11}) we analytically obtain the corresponding mean Uhlmann curvature
matrix, 
\begin{equation}
D=\left( 
\begin{array}{cc}
0 & \Delta _{1}\Delta _{2} \\ 
-\Delta _{1}\Delta _{2} & 0%
\end{array}%
\right) ,  \label{29}
\end{equation}%
where 
\begin{eqnarray}
\Delta _{1} &=&e^{-2\tau \coth (\pi a)}\sin \theta ,  \notag \\
\Delta _{2} &=&1+(1-e^{\tau \coth (\pi a)})\tanh (\pi a)\cos \theta .
\label{30}
\end{eqnarray}

These results demonstrate that the SLD-CRB remains unattainable, even in the
asymptotic limit of measurements performed on an asymptotically large number
of copies of the two-level atom system. Then, by exploiting Eqs. (\ref{5}), (%
\ref{6}) and (\ref{7}), we obtain the SLD-CRB $C_{(\theta ,\phi )}^{S}$ and
the RLD-CRB $C_{(\theta ,\phi )}^{R}$, 
\begin{eqnarray}
C_{(\theta ,\phi )}^{S} &=&\left( \csc ^{2}\theta +\Theta /\Lambda \right)
e^{\tau \coth (\pi a)},  \notag \\
C_{(\theta ,\phi )}^{R} &=&\Xi \Theta /\Lambda +2\sqrt{\Upsilon ^{2}/\Lambda
^{2}},  \label{31}
\end{eqnarray}%
where we have set%
\begin{eqnarray}
\Lambda &=&\Lambda _{1}+2\Lambda _{2},  \notag \\
\Theta &=&\Theta _{1}+\Theta _{2},  \notag \\
\Upsilon &=&\Upsilon _{1}\Upsilon _{2}\Theta ,  \notag \\
\Xi &=&1+e^{\tau \coth (\pi a)}\csc ^{2}\theta ,  \label{32}
\end{eqnarray}%
with%
\begin{eqnarray}
\Lambda _{1} &=&[3+\cos (2\theta )]\cosh (2\pi a),  \notag \\
\Lambda _{2} &=&2e^{\tau \coth (\pi a)}\cos ^{2}\theta +\sin ^{2}\theta
+2\sinh (2\pi a)\cos \theta ,  \notag \\
\Theta _{1} &=&4e^{\tau \coth (\pi a)}+2\cos (2\theta )\cosh ^{2}(\pi a), 
\notag \\
\Theta _{2} &=&3\cosh (2\pi a)+4\sinh (2\pi a)\cos \theta -1,  \notag \\
\Upsilon _{1} &=&\coth (\pi a)+(1-e^{\tau \coth (\pi a)})\cos \theta , 
\notag \\
\Upsilon _{2} &=&\tanh (\pi a)\csc \theta .  \label{33}
\end{eqnarray}

\begin{figure}[tbp]
\centering
\makebox[\linewidth]{\includegraphics[width=0.5\linewidth]{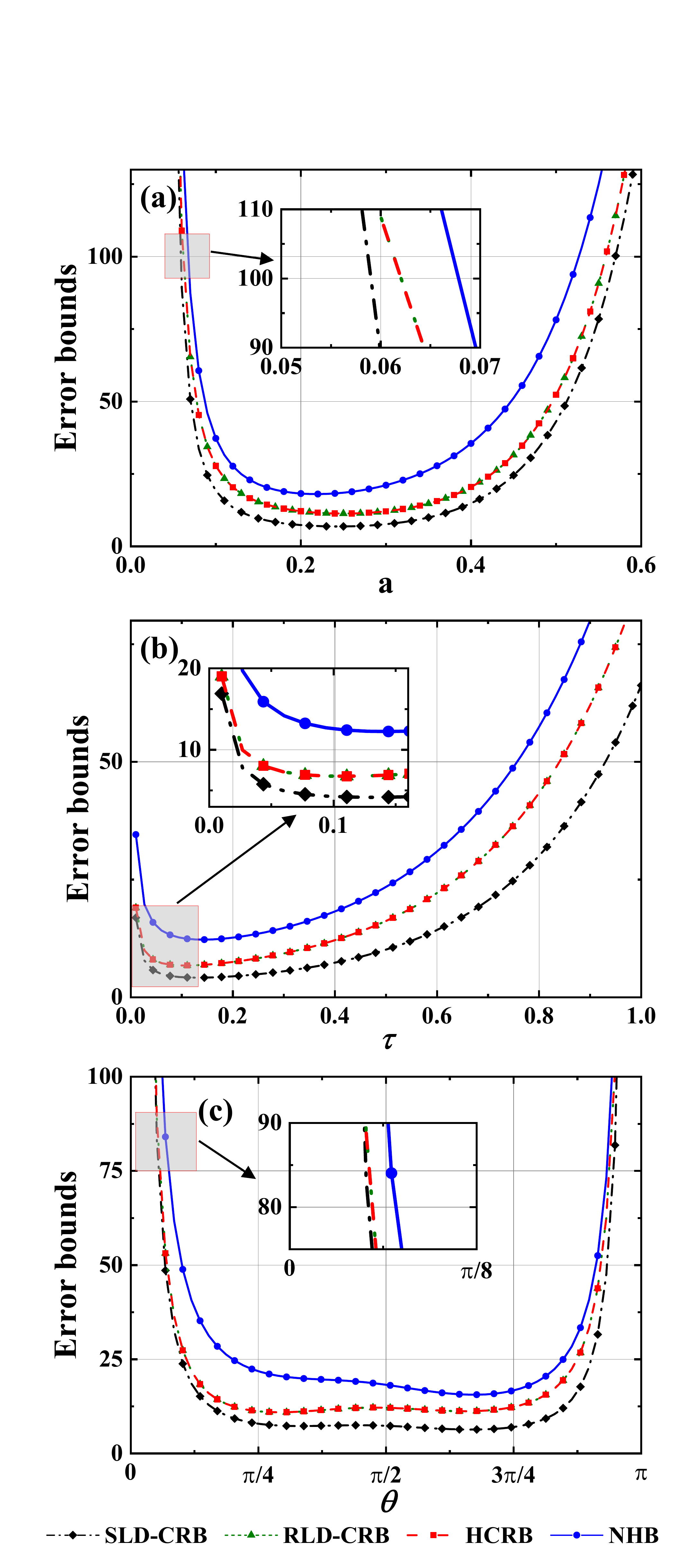}}

\caption{Error bounds as a function of (a) the inverse of acceleration $a$ 
with $\theta = \pi/2$ and $\tau = 0.4$, (b) the proper time $\tau$ 
with $\theta = \pi/2$ and $a = 0.2$, (c) the weight parameter $\theta$ 
with $\tau = 0.4$ and $a = 0.2$.}
\label{Fig2}
\end{figure}

Typically, the calculation of the HCRB $C_{(\theta ,\phi )}^{H}$ and NB $%
C_{(\theta ,\phi )}^{N}$ requires to solve a minimization problem formulated
as a semidefinite program, which is computationally non-trivial. However,
for two-parameter estimation in a single-qubit system, the corresponding
HCRB $C_{(\theta ,\phi )}^{H}$ and NB $C_{(\theta ,\phi )}^{N}$ can be
obtained through analytic expressions Refs. \cite{55,76}, 
\begin{eqnarray}
C_{(\theta ,\phi )}^{H} &\text{=}&\left\{ 
\begin{array}{c}
C_{(\theta ,\phi )}^{R},\text{ }C_{(\theta ,\phi )}^{R}\geq \frac{C_{(\theta
,\phi )}^{S}+C_{(\theta ,\phi )}^{Z}}{2} \\ 
C_{(\theta ,\phi )}^{R}+S_{(\theta ,\phi )},\text{ }C_{(\theta ,\phi )}^{R}<%
\frac{C_{(\theta ,\phi )}^{S}+C_{(\theta ,\phi )}^{Z}}{2}%
\end{array}%
\right. ,  \notag \\
C_{(\theta ,\phi )}^{N} &\text{=}&C_{(\theta ,\phi )}^{S}+2\sqrt{\Theta _{3}}%
,  \label{34}
\end{eqnarray}%
where 
\begin{eqnarray}
S_{(\theta ,\phi )} &=&\frac{\left[ \frac{1}{2}(C_{(\theta ,\phi
)}^{S}+C_{(\theta ,\phi )}^{Z})-C_{(\theta ,\phi )}^{R}\right] ^{2}}{%
C_{(\theta ,\phi )}^{Z}-C_{(\theta ,\phi )}^{R}},  \notag \\
C_{(\theta ,\phi )}^{Z} &=&C_{(\theta ,\phi )}^{S}+2\sqrt{\left. \Theta
^{2}\Lambda _{3}^{2}\csc ^{2}\theta \right/ \Lambda ^{2}},  \notag \\
\Lambda _{3} &=&(e^{\tau \coth (\pi a)}-1)\tanh (\pi a)\cos \theta -1, 
\notag \\
\Theta _{3} &=&\frac{\Theta e^{2\tau \coth (\pi a)}\csc ^{2}\theta }{\Lambda 
}.  \label{35}
\end{eqnarray}%
\ \ \ \ \ \ \ Fig. 1 presents the numerical results via SDP, which compare
SLD-CRB, RLD-CRB, HCRB and NB as functions of the relevant physical
parameters. Notably, the NB consistently yields the largest values among all
the bounds, confirming its role as the tightest achievable precision limit.
In Fig. 1(a), it is evident that all error bounds decrease monotonically
with the increasing inverse of acceleration $a.$\ Moreover, the RLD-CRB and
HCRB are nearly identical in numerical value, indicating that both provide
an asymptotically tight precision limit. In Fig. 1(b), all error bounds
clearly increase with the increase of the proper time $\tau .$\ For $\tau
<1.147,$\ the RLD-CRB and HCRB are almost equal and exceed the SLD-CRB,
suggesting that both RLD-CRB and HCRB serve as asymptotically tight
precision limits, while the SLD-CRB exhibits the poorest tightness. In
contrast, for $\tau >1.147,$\ the SLD-CRB and HCRB become nearly identical
and surpass the RLD-CRB, implying that both SLD-CRB and HCRB provide an
asymptotically tight precision limit, with the RLD-CRB being the least
tight. In Fig. 1(c), all error bounds exhibit a non-monotonic behavior,
first decreasing and then increasing with the weight parameter $\theta $.
Specifically, for $\theta <0.593,$\ the SLD-CRB, RLD-CRB, and HCRB are
nearly equal. In the range $0.593<\theta <1.443,$\ the SLD-CRB and HCRB are
almost identical and larger than the RLD-CRB. For $1.443<\theta <2.823,$\
the RLD-CRB and HCRB are nearly the same and exceed the SLD-CRB. For $\theta
>2.823,$\ the RLD-CRB, HCRB and NB are almost equal, indicating that these
error bounds can provide tight precision limits.

Next, we analyze a three-parameter estimation problem involving the weight
parameter $\theta $, phase parameter $\phi $\ and inverse of acceleration $a$%
\ in the two-level atom system. Due to the complexity of the analytical
results, we focus on numerical comparisons among the SLD-CRB, RLD-CRB, HCRB
and NHB as functions of the relevant physical parameters, as shown in Fig.
2. It is clearly observed that the NHB consistently produces the largest
values, confirming its position as the tightest error bound. The RLD-CRB and
HCRB are numerically equal and both exceed the SLD-CRB, indicating that the
RLD-CRB and HCRB each provide an asymptotically tight precision limit,
whereas the SLD-CRB shows the weakest tightness. Furthermore, these bounds
follow a non-monotonic trend, initially decreasing and then increasing with
the variations in the inverse of acceleration $a,$\ proper time $\tau $\ and
weight parameter $\theta .$

\subsection{Multiparameter quantum estimation with a boundary}

We introduce a boundary at $z$=$0$ and analyze a uniformly accelerated atom
moving in the $x-y$ plane at a distance $z$ from the boundary \cite{28,35,75}%
. In this scenario, the two-point correlation function can be described as 
\cite{28,35,75}, 
\begin{equation}
G^{+}(x,x^{\prime })=G^{+}(x,x^{\prime })_{0}+G^{+}(x,x^{\prime })_{b},
\label{36}
\end{equation}%
where $G^{+}(x,x^{\prime })_{0}$ is the two-point correlation function
without boundary that can be obtained from Eq. (\ref{26}), the second term 
\begin{equation}
G^{+}(x,x^{\prime })_{b}=-\frac{1/(4\pi ^{2})}{L-T},  \label{37}
\end{equation}%
accounts for the correction induced by the presence of the boundary, where $%
L $=$(x-x^{\prime })^{2}+(y-y^{\prime })^{2}+(z+z^{\prime })^{2}$ and $T$=$%
(t-t^{\prime }-i\epsilon )^{2}$. Using the trajectory of the two-level atom
from Eq. (\ref{25}), we derive the specific form of the two-point
correlation function \cite{28,75}, 
\begin{equation}
G^{+}(x,x^{\prime })=-\frac{a^{2}}{16\pi ^{2}}\left[ \frac{1}{S}-\frac{1}{%
S-a^{2}z^{2}}\right] ,  \label{38}
\end{equation}%
where $S$=$\sinh ^{2}(a\Delta \tau /2-i\epsilon ).$%

\begin{figure}[tbp]
\centering
\includegraphics[width=0.5\linewidth]{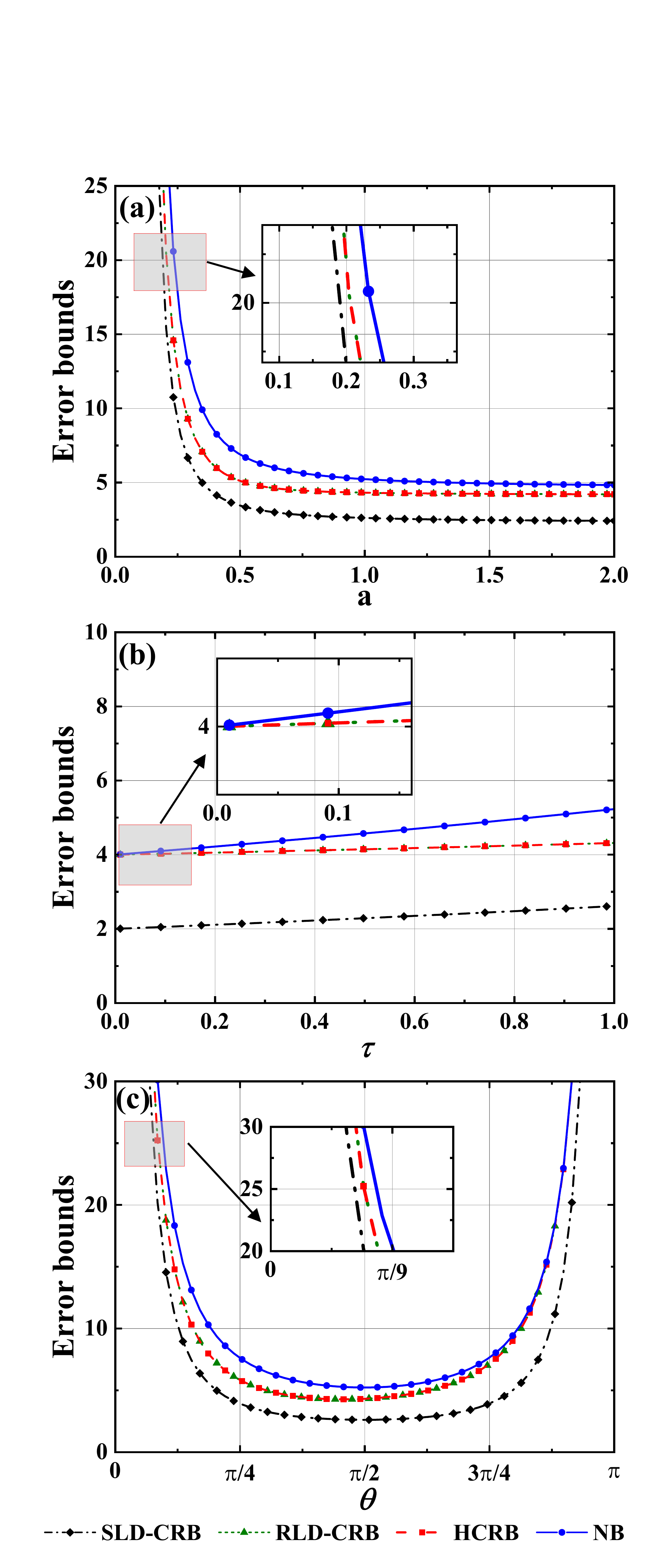}
\caption{Error bounds as a function of (a) the inverse of acceleration $a$ 
with $\theta = \pi/2$, $\tau = 1$ and $z = 0.5$, (b) the proper time $\tau$ 
with $\theta = \pi/2$, $a = 1$ and $z = 0.5$, (c) the weight parameter $\theta$ 
with $\tau = 1$, $a = 1$ and $z = 0.5$.}
\label{Fig3}
\end{figure}

According to Eq. (\ref{22}), we also obtain the Fourier transformation of
the two-point correlation function \cite{28,75}, 
\begin{equation}
G(\omega _{0})_{b}=G(\omega _{0})_{0}\left\{ 1-\frac{\sin \left[ \frac{%
2\omega _{0}}{a}arc\sinh (az)\right] }{2z\omega _{0}\sqrt{1+a^{2}z^{2}}}%
\right\} ,  \label{39}
\end{equation}%
where $G(\omega _{0})_{0}$ is given by Eq. (\ref{27}), which enables us to
determine the coefficients for the Kossakowski matrix based on Eq. (\ref{21}%
), 
\begin{eqnarray}
A_{b} &=&A_{0}\left\{ 1-\frac{\sin \left[ \frac{2\omega _{0}}{a}arc\sinh (az)%
\right] }{2z\omega _{0}\sqrt{1+a^{2}z^{2}}}\right\} ,  \notag \\
B_{b} &=&B_{0}\left\{ 1-\frac{\sin \left[ \frac{2\omega _{0}}{a}arc\sinh (az)%
\right] }{2z\omega _{0}\sqrt{1+a^{2}z^{2}}}\right\} ,  \label{40}
\end{eqnarray}%
where $A_{0}$ and $B_{0}$ are defined by Eq. (\ref{28}). 

\begin{figure}[tbp]
\centering
\makebox[\linewidth]{\includegraphics[width=0.5\linewidth]{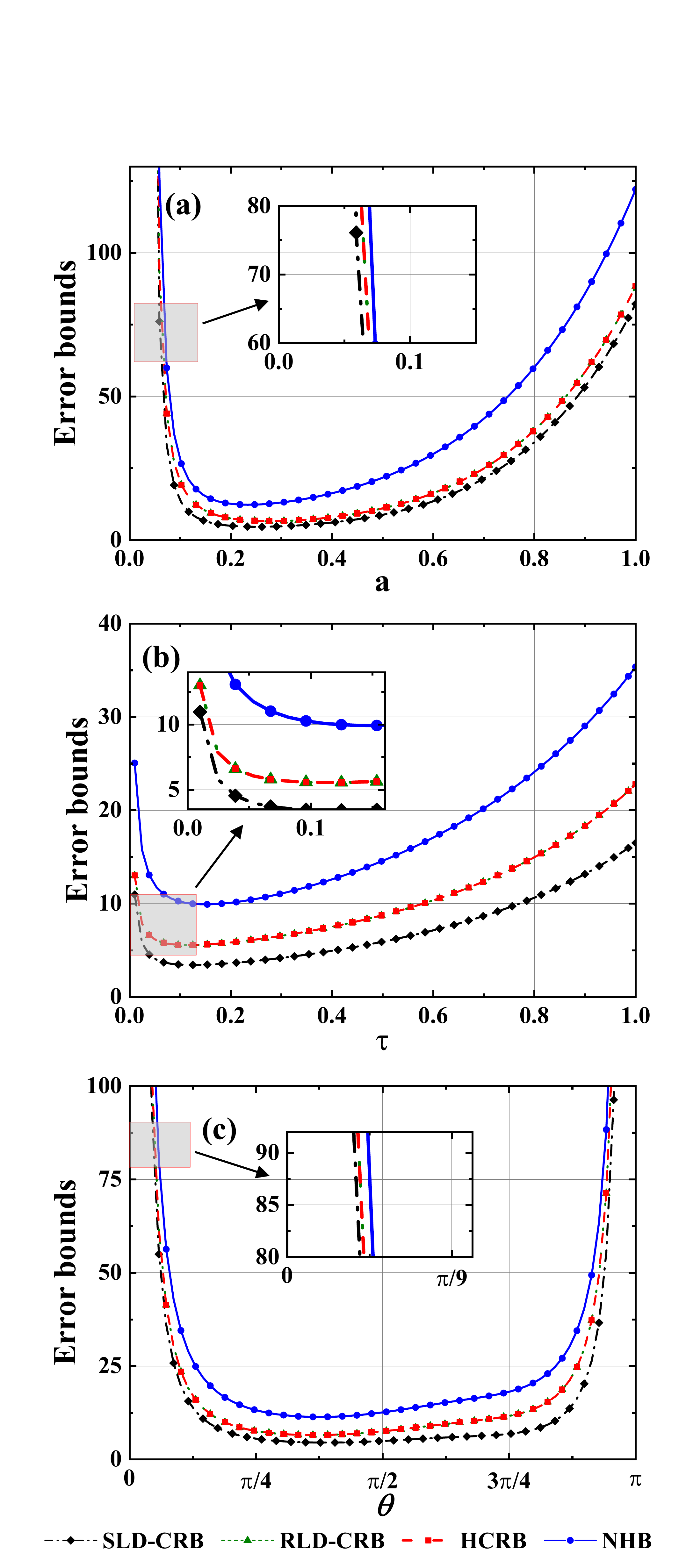}}

\caption{Error bounds as a function of (a) the inverse of acceleration $a$ 
with $\theta = \pi/2$, $\tau = 0.4$ and $z = 0.5$, (b) the proper time $\tau$ 
with $\theta = \pi/2$, $a = 0.2$ and $z = 0.5$, (c) the weight parameter $\theta$ 
with $\tau = 0.4$, $a = 0.2$ and $z = 0.5$.}
\label{Fig4}
\end{figure}

Further, substituting Eq. (\ref{40}) into Eq. (\ref{23}), we get the
time-dependent reduced density matrix of the uniformly accelerated two-level
atom system with a boundary. In the following discussion and analysis, we
utilize the transformations $\tau \rightarrow \tilde{\tau}$=$\Gamma _{0}\tau 
$, $a\rightarrow $ $\tilde{a}$=$\omega _{0}/a,$ and $z\rightarrow \tilde{z}$=%
$z\omega _{0}.$ For simplicity, $\tilde{\tau}$, $\tilde{a}$ and $\tilde{z}$
will be denoted as $\tau $, $a$ and $z$, respectively. Similarly, we
initially take into account a two-parameter estimation that pertains to the
initial weight parameter $\theta $ and phase parameter $\phi $ of the
two-level atom system$.$ Owing to the complexity of the analytical outcomes,
we concentrate on the numerical comparisons of the SLD-CRB, RLD-CRB, HCRB
and NB as functions of the relevant physical parameters, as depicted in Fig.
3. Generally, HCRB represents an asymptotically tight precision limit that
is achievable through collective measurements, while NB constitutes a tight
precision limit attainable via single-copy measurements. In comparison with
Fig. 1, we observe that at a fixed value of $z=0.5$, the introduction of the
boundary reduces the numerical values of both HCRB and NB, indicating an
improvement in the attainable estimation precision. In Fig. 3(a), all error
bounds exhibit a monotonic decrease and asymptotically approach a non-zero
value as the inverse of acceleration $a.$\ Moreover, the RLD-CRB and HCRB
are numerically identical, indicating that both provide an asymptotically
tight precision limit. In Fig. 3(b), all error bounds show a monotonic
increase with the proper time $\tau $. In the regime where $\tau <0.172,$\
RLD-CRB, HCRB, and NB are nearly identical, suggesting that these bounds
provide tight precision limits. In Fig. 3(c), all error bounds display a
non-monotonic behavior, first decreasing and then increasing with the weight
parameter $\theta $. For $\theta >2.662,$\ RLD-CRB, HCRB and NB are almost
equal, indicating that these error bounds can provide tight precision limits.

Subsequently, we consider a three-parameter estimation problem including the
weight parameter $\theta $, phase parameter $\phi $ and inverse of
acceleration $a$ in the two-level atom system with a boundary. As
illustrated in Fig. 4, the SLD-CRB, RLD-CRB, HCRB and NHB are evaluated as
functions of the relevant physical parameters. Comparative analysis with
Fig. 2 reveals that the introduction of the boundary systematically reduces
the numerical values of both HCRB and NHB. This also implies that the
corresponding estimation precision has been enhanced. Notably, the NHB
consistently demonstrates the largest values across all parameter
configurations, thereby confirming its established status as the ultimate
achievable precision limit in this scenario.\ The RLD-CRB and HCRB are
numerically equivalent and both surpass the SLD-CRB, indicating that the
RLD-CRB and HCRB each serve as an asymptotically tight precision limit,
whereas the SLD-CRB exhibits the weakest tightness. Furthermore, as the
inverse of acceleration $a,$ proper time $\tau $ and weight parameter $%
\theta $ increase, all bounds exhibit a characteristic non-monotonic
dependence on the parameter variations, displaying initial decline followed
by later upward trend.

\section{Conclusions}

In summary, we have conducted a comprehensive analysis on multiparameter
quantum estimation for a uniformly accelerated Unruh-DeWitt detector
interacting with a vacuum scalar field in both bounded and unbounded
Minkowski vacuum. In the unbounded scenario, we have initially investigated
a two-parameter estimation problem involving the initial weight parameter
and phase parameter of the Unruh-DeWitt detector. We have derived analytical
expressions for the SLD-CRB, RLD-CRB, HCRB and NB, and numerically solved
these bounds using the SDP. Our results demonstrate that the NB yields the
tightest error bound among all bounds, consistent with the general hierarchy
of multiparameter quantum estimation. Notably, while these error bounds vary
monotonically with the inverse of acceleration and proper time, they exhibit
non-monotonic behavior with respect to the weight parameter. More
importantly, we observe a significant competition and alternation in
tightness between the SLD-CRB and RLD-CRB, particularly as proper time and
the inverse of acceleration vary. This implies a transition in the physical
mechanisms governing measurement precision across different dynamical
evolution stages or parameter configurations. Consequently, relying on a
single SLD-CRB or RLD-CRB is often insufficient; the HCRB and NB, however,
consistently provides an asymptotically tight precision limit, highlighting
the necessity of employing these tighter bounds in multiparameter quantum
estimation. Subsequently, we have explored a three-parameter estimation
problem involving the weight parameter, phase parameter and inverse of
acceleration in the Unruh-DeWitt detector. Our numerical findings reveal
that the NHB consistently produces the largest values, confirming its status
as the tightest bound. The RLD-CRB and HCRB are numerically identical and
both exceed the SLD-CRB, indicating that they offer asymptotically tight
precision limits, whereas the SLD-CRB shows the weakest tightness. In the
case with a boundary, we have also examined both two-parameter and
three-parameter estimation in the Unruh-DeWitt detector. While observing
similar trends to the unbounded case, we find that the introduction of the
boundary reduces the numerical values of HCRB, NB and NHB, thereby an
improvement in the attainable estimation precision.

\begin{acknowledgments}
We sincerely thank Marco G. Genoni, Francesco Albarelli, and Simon K. Yung
for helpful discussions. This work was supported by the National Key
Research and Development Program of China (Grants Nos. 2021YFA1400900,
2021YFA0718300, and 2021YFA1400243) and NSFC (Grants Nos. 12074105,
12104135, 12404377, and No. 61835013). Wei Ye is supported by Jiangxi
Provincial Natural Science Foundation (20232BAB211032, 20252BAC240169) and
the Scientific Research Startup Foundation (EA202204230).
\end{acknowledgments}

\raggedright
\textbf{Data Availability Statement.} This article has no associated data or
the data will not be deposited.

\textbf{Code Availability Statement.} This article has no associated code or
the code will not be deposited.

\textbf{Open Access.} This article is distributed under the terms of the
Creative Commons Attri bution License (CC-BY4.0), which permits any use,
distribution and reproduction in any medium, provided the original author(s)
and source are credited.


\begin{thebibliography}{99}

\bibitem{1} 
S. Pirandola, B. R. Bardhan, T. Gehring, C. Weedbrook and S. Lloyd, 
\textit{Advances in photonic quantum sensing}, \href{https://arxiv.org/abs/1811.01969}
{Nat. Photon. \textbf{12} (2018) 724} [\href{https://arxiv.org/abs/1811.01969}{{arXiv:1811.01969}}][\href{https://arxiv.org/abs/1811.01969}{{inSPIRE}}].

\bibitem{2} V. Giovannetti, S. Lloyd and L. Maccone, Advances in quantum metrology,\href{http://hdl.handle.net/1721.1/78939}{Nat. Photonics \textbf{5} (2011) 222}
[\href{http://hdl.handle.net/1721.1/78939}{{arXiv:2411.03850v2}}]
[\href{http://hdl.handle.net/1721.1/78939}{{inSPIRE}}].

\bibitem{3} C. Mukhopadhyay, V. Montenegro and A. Bayat, Current trends in
global quantum metrology, \href{https://doi.org/10.48550/arXiv.2411.03850}{J. Phys. A: Math. Theor. \textbf{58} (2025) 063001}
[\href{https://doi.org/10.48550/arXiv.2411.03850}{{arXiv:2411.03850v2}}]
[\href{https://doi.org/10.48550/arXiv.2411.03850}{{inSPIRE}}].

\bibitem{4} M. A. Rodr\'{\i}uez-Garc\'{\i}a, R. L. de Matos Filho and P.
Barberis-Blostein, Usefulness of quantum entanglement for enhancing
precision in frequency estimation, \href{https://doi.org/10.48550/arXiv.2405.06548}{Phys. Rev. Res. \textbf{6} (2024) 043230}
[\href{https://doi.org/10.48550/arXiv.2405.06548}{{arXiv:2405.06548}}]
[\href{https://doi.org/10.48550/arXiv.2405.06548}{{inSPIRE}}].

\bibitem{5} R. Demkowicz-Dobrza\'{n}ski and L. Maccone, Using entanglement
against noise in quantum metrology,\href{https://doi.org/10.1103/PhysRevLett.113.250801}{ Phys. Rev. Lett. \textbf{113 }(2014)
250801 }
[\href{https://doi.org/10.1103/PhysRevLett.113.250801}{arXiv:1407.2934}][\href{https://doi.org/10.1103/PhysRevLett.113.250801}{inSPIRE}].

\bibitem{6} H. Zhang, W. Ye, Z. Liao and X. Wang, Quantum superresolution
for imaging two pointlike entangled photon sources, \href{ https://doi.org/10.1103/PhysRevA.108.033713}{Phys. Rev. A \textbf{108 
}(2023) 033713}
[\href{ https://doi.org/10.1103/PhysRevA.108.033713}{arXiv:2306.10238v1}]
[\href{ https://doi.org/10.1103/PhysRevA.108.033713}{inSPIRE}].

\bibitem{7} L. Pezz\`{e}, A. Smerzi, M. K. Oberthaler, R. Schmied and P.
Treutlein, Quantum metrology with nonclassical states of atomic ensembles,
\href{https://doi.org/10.1103/PhysRevLett.134.180801}{Rev. Mod. Phys. \textbf{90 }(2018) 035005 }
[\href{https://doi.org/10.1103/PhysRevLett.134.180801}{arXiv:1609.01609}]
[\href{https://doi.org/10.1103/PhysRevLett.134.180801}{inSPIRE}].

\bibitem{8} Q. R. Rahman, I. Kladari\'{c}, M. Kern, L. Lachman, Y. Chu, R.
Filip and M. Fadel, Genuine quantum non-Gaussianity and metrological
sensitivity of Fock States prepared in a mechanical resonator,\href{https://doi.org/10.1103/PhysRevLett.134.180801}{Phys. Rev.Lett. \textbf{134 }(2025) 180801 }
[\href{https://doi.org/10.1103/PhysRevLett.134.180801}{arXiv:2412.20971v3}][\href{https://doi.org/10.1103/PhysRevLett.134.180801}{inSPIRE}].

\bibitem{9} S. Elghaayda, A. Ali, M. Y. Abd-Rabbou, M. Mansour and S.
Al-Kuwari, Quantum correlations and metrological advantage among
Unruh--DeWitt detectors in de Sitter spacetime,\href{https://doi.org/10.48550/arXiv.2412.07425}{Eur. Phys. J. C \textbf{85 }%
(2025) 447 }
[\href{https://doi.org/10.48550/arXiv.2412.07425}{arXiv:2412.07425}][\href{https://doi.org/10.48550/arXiv.2412.07425}{inSPIRE}].

\bibitem{10} J. Sahota and N. Quesada, Quantum correlations in optical
metrology: Heisenberg-limited phase estimation without mode entanglement,\href{https://doi.org/10.1103/PhysRevA.91.013808}{
Phys. Rev. A \textbf{91 }(2015) 013808}
[\href{https://doi.org/10.1103/PhysRevA.91.013808}{arXiv:1404.7110v2}][\href{https://doi.org/10.1103/PhysRevA.91.013808}{inSPIRE}].

\bibitem{11} M. A. Ciampini, N. Spagnolo, C. Vitelli, L. Pezz\`{e}, A.
Smerzi and F. Sciarrino, Quantum-enhanced multiparameter estimation in
multiarm interferometers, \href{https://doi.org/10.48550/arXiv.1507.07814}{Sci. Rep. \textbf{6 }(2016) 28881}
[\href{https://doi.org/10.48550/arXiv.1507.07814}{arXiv:1507.07814}]
[\href{https://doi.org/10.48550/arXiv.1507.07814}{inSPIRE}].

\bibitem{12} J. S. Sidhu and P. Kok, Geometric perspective on quantum
parameter estimation,\href{https://doi.org/10.48550/arXiv.1907.06628}{ AVS Quantum Sci. \textbf{2 }(2020) 014701}
[\href{https://doi.org/10.48550/arXiv.1907.06628}{arXiv:1907.06628}]
[\href{https://doi.org/10.48550/arXiv.1907.06628}{inSPIRE}].

\bibitem{13} R. Demkowicz-Dobrza\'{n}ski, M. Jarzyna and J. Ko\l ody\'{n}%
ski, Quantum limits in optical interferometry, \href{https://doi.org/10.48550/arXiv.1405.7703}{Prog. Opt. \textbf{60 }(2015)345} 
[\href{https://doi.org/10.48550/arXiv.1405.7703}{arXiv:1405.7703}][\href{https://doi.org/10.48550/arXiv.1405.7703}{inSPIRE}].

\bibitem{14} M. G. A. Paris, Quantum estimation for quantum technology, Int.\href{https://api.semanticscholar.org/CorpusID:2365312}{
J. Quantum Inf. \textbf{7 }(2009) 125 }
[\href{https://api.semanticscholar.org/CorpusID:2365312}{arXiv:0804.2981}][\href{https://api.semanticscholar.org/CorpusID:2365312}{inSPIRE}].

\bibitem{15} M. Gessner and A. Smerzi, Hierarchies of frequentist bounds for
quantum metrology: From Cram\'{e}r-Rao to Barankin,\href{https://doi.org/10.1103/PhysRevLett.130.260801}{ Phys. Rev. Lett. \textbf{%
130 }(2023) 260801 }
[\href{https://doi.org/10.1103/PhysRevLett.130.260801}{arXiv:2303.06108}][\href{https://doi.org/10.1103/PhysRevLett.130.260801}{inSPIRE}].

\bibitem{16} V. Gebhart, M. Gessner and A. Smerzi, Fundamental bounds for
parameter estimation with few measurements,\href{https://doi.org/10.48550/arXiv.2402.14495}{ Phys. Rev. Res. \textbf{6 }%
(2024) 043261}
[\href{https://doi.org/10.48550/arXiv.2402.14495}{arXiv:2402.14495}]
[\href{https://doi.org/10.48550/arXiv.2402.14495}{inSPIRE}].

\bibitem{17} V. Montenegro, C. Mukhopadhyay, R. Yousefjani, S. Sarkar, U.
Mishra, M. G. A. Paris and A. Bayat, Review: quantum metrology and sensing
with many-body systems,\href{https://doi.org/10.1016/j.physrep.2025.05.005}{ Phys. Rep. \textbf{1134 }(2025) 1-62}
[\href{https://doi.org/10.1016/j.physrep.2025.05.005}{arXiv:2408.15323}]
[\href{https://doi.org/10.1016/j.physrep.2025.05.005}{inSPIRE}].

\bibitem{18} M. Reichert, Q. Zhuang and M. Sanz, Heisenberg-limited quantum
lidar for joint range and velocity estimation,\href{https://doi.org/10.1103/PhysRevLett.133.130801}{ Phys. Rev. Lett. \textbf{133 }%
(2024) 130801 }
[\href{https://doi.org/10.1103/PhysRevLett.133.130801}{arXiv:2311.14546v2}]
[\href{https://doi.org/10.1103/PhysRevLett.133.130801}{inSPIRE}].

\bibitem{19} G. W. Qian, X. Q. Xu, S. A. Zhu, C. R. Xu, F. Gao, V. V.
Yakovlev, X. Liu, S. Y. Zhu and D. W. Wang, Quantum induced coherence light
detection and ranging, \href{https://doi.org/10.1103/PhysRevLett.131.033603}{Phys. Rev. Lett. \textbf{131 }(2023) 033603}
[\href{https://doi.org/10.1103/PhysRevLett.131.033603}{arXiv:2212.12924}]
[\href{https://doi.org/10.1103/PhysRevLett.131.033603}{inSPIRE}].

\bibitem{20} E. T. Khabiboulline, J. Borregaard, K. De Greve and M. D.
Lukin, Quantum-assisted telescope arrays, \href{https://doi.org/10.1103/PhysRevA.100.022316}{Phys. Rev. A \textbf{100 }(2019)
022316}
[\href{https://doi.org/10.1103/PhysRevA.100.022316}{arXiv:1809.03396}]
[\href{https://doi.org/10.1103/PhysRevA.100.022316}{inSPIRE}].

\bibitem{21} D. Gottesman, T. Jennewein and S. Croke, Longer-baseline
telescopes using quantum repeaters,\href{https://doi.org/10.1103/PhysRevLett.109.070503}{ Phys. Rev. Lett. \textbf{109 }(2012)
070503 }
[\href{https://doi.org/10.1103/PhysRevLett.109.070503}{arXiv:1107.2939}]
[\href{https://doi.org/10.1103/PhysRevLett.109.070503}{inSPIRE}].

\bibitem{22} H. R. Jahromi, S. E. A. Mamaghani and R. L. Franco,
Relativistic quantum thermometry through a moving sensor,\href{https://doi.org/10.1016/j.aop.2022.169172}{ Ann. Phys. \textbf{%
448 }(2023) 169172}
[\href{https://doi.org/10.1016/j.aop.2022.169172}{arXiv:2208.04431}]
[\href{https://doi.org/10.1016/j.aop.2022.169172}{inSPIRE}].

\bibitem{23} J. Rubio, J. Anders and L. A. Correa, Global quantum
thermometry,\href{https://doi.org/10.1103/PhysRevLett.127.190402}{ Phys. Rev. Lett. \textbf{127 }(2021) 190402 }
[\href{https://doi.org/10.1103/PhysRevLett.127.190402}{arXiv:2011.13018}]
[\href{https://doi.org/10.1103/PhysRevLett.127.190402}{inSPIRE}].

\bibitem{24} S. K. Chang, Y. Yan, L. Wang, W. Ye, X. Rao, H. Zhang, L.
Huang, M. Luo, Y. Chen, Q. Ma and S. Gao, Global quantum thermometry based
on the optimal biased bound,\href{https://doi.org/10.1103/PhysRevResearch.6.043171}{ Phys. Rev. Res. \textbf{6 }(2024) 043171}
[\href{https://doi.org/10.1103/PhysRevResearch.6.043171}{arXiv:2305.08397}]
[\href{https://doi.org/10.1103/PhysRevResearch.6.043171}{inSPIRE}].

\bibitem{25} M. Ahmadi, D. E. Bruschi and I. Fuentes, Quantum metrology for
relativistic quantum fields, \href{https://doi.org/10.1103/PhysRevD.89.065028}{Phys. Rev. D \textbf{89 }(2014) 065028}
[\href{https://doi.org/10.1103/PhysRevD.89.065028}{arXiv:1312.5707}]
[\href{https://doi.org/10.1103/PhysRevD.89.065028}{inSPIRE}].

\bibitem{26} H. Du and R. B. Mann, Fisher information as a probe of
spacetime structure: relativistic quantum metrology in (A)dS,\href{https://doi.org/10.1007/jhep05(2021)112}{ J. High Energ.
Phys \textbf{05 }(2021) 112 }
[\href{https://doi.org/10.1007/jhep05(2021)112}{arXiv:2012.08557}]
[\href{https://doi.org/10.1007/jhep05(2021)112}{inSPIRE}].

\bibitem{27} E. Pattersona and R. B. Manna, Fisher information of a black
hole spacetime, \href{https://doi.org/10.1007/jhep06(2023)214}{J. High Energ. Phys \textbf{06 }(2023)\textbf{\ }214}
[\href{https://doi.org/10.1007/jhep06(2023)214}{arXiv:2207.12226}]
[\href{https://doi.org/10.1007/jhep06(2023)214}{inSPIRE}].

\bibitem{28} Z. Zhao, Q. Pan and J. Jing, Quantum estimation of acceleration
and temperature in open quantum system,\href{https://doi.org/10.1103/PhysRevD.101.056014}{ Phys. Rev. D \textbf{101 }(2020)
056014}
[\href{https://doi.org/10.1103/PhysRevD.101.056014}{arXiv:2007.13389}]
[\href{https://doi.org/10.1103/PhysRevD.101.056014}{inSPIRE}].

\bibitem{29} L. Chen and J. Feng, Quantum Fisher information of a cosmic
qubit undergoing non-Markovian de Sitter evolution,\href{https://doi.org/10.1007/JHEP06%282025%29029}{ J. High Energ. Phys  \textbf{06 }(2025) 029 }
[\href{https://doi.org/10.1007/JHEP06%282025%29029}{arXiv:2411.11490}]
[\href{https://doi.org/10.1007/JHEP06%282025%29029}{inSPIRE}].

\bibitem{30} J. Feng and J. Zhang, Quantum Fisher information as a probe for
Unruh thermality, \href{https://doi.org/10.1016/j.physletb.2022.136992}{Phys. Lett. B \textbf{827 }(2022) 136992}
[\href{https://doi.org/10.1016/j.physletb.2022.136992}{arXiv:2111.00277}]
[\href{https://doi.org/10.1016/j.physletb.2022.136992}{inSPIRE}].

\bibitem{31} Z. Tian, J. Wang, J. Jing and H. Fan, Relativistic quantum metrology in open system dynamics, \href{https://doi.org/10.1038/srep07946}{Sci. Rep. \textbf{5} (2015) 7946}
[\href{https://doi.org/10.1038/srep07946}{arXiv:1501.06676}]    
[\href{https://doi.org/10.1038/srep07946}{inSPIRE}].

\bibitem{32} M. Aspachs, G. Adesso and I. Fuentes, Optimal quantum
estimation of the Unruh-Hawking effect,\href{ https://doi.org/10.1103/PhysRevLett.105.151301}{ Phys. Rev. Lett. \textbf{105 }(2010)
151301}
[\href{ https://doi.org/10.1103/PhysRevLett.105.151301}{arXiv:1007.0389}]
[\href{ https://doi.org/10.1103/PhysRevLett.105.151301}{inSPIRE}].

\bibitem{33} X. Liu, J. Jing, Z. Tian and W. Yao, Does relativistic motion
always degrade quantum Fisher information?
,\href{https://doi.org/10.1103/PhysRevD.103.125025}{ Phys. Rev. D \textbf{103 }(2021)125025 }
[\href{https://doi.org/10.1103/PhysRevD.103.125025}{arXiv:2205.08725}][\href{https://doi.org/10.1103/PhysRevD.103.125025}{inSPIRE}].

\bibitem{34} H. Wang, J. Zhang and H. Yu, Quantum parameter estimation for
detectors in constantly accelerated motion ,\href{https://doi.org/10.1016/j.aop.2016.04.021}{Phys. Rev. D \textbf{112 }(2025)045006 }
[\href{https://doi.org/10.1016/j.aop.2016.04.021}{arXiv:2503.11016}][\href{https://doi.org/10.1016/j.aop.2016.04.021}{inSPIRE}].

\bibitem{35} Z. Zhao, S. Zhang, Q. Pan and J. Jing, Estimation precision of
parameter associated with Unruh-like effect, ,\href{https://doi.org/10.1016/j.nuclphysb.2021.115408}{Nucl. Phys. B \textbf{967 }%
(2021) 115408 }[\href{https://doi.org/10.1016/j.nuclphysb.2021.115408}{arXiv:2007.14794}][\href{https://doi.org/10.1016/j.nuclphysb.2021.115408}{inSPIRE}].

\bibitem{36} F. Albarelli, M. Barbieri, M. G. Genoni and I. Gianani, A
perspective on multiparameter quantum metrology: From theoretical tools to
applications in quantum imaging
,\href{https://doi.org/10.1016/j.physleta.2020.126311}{ Phys. Lett. A \textbf{384 }(2020) 126311}[\href{https://doi.org/10.1016/j.physleta.2020.126311}{arXiv:1911.12067}][\href{https://doi.org/10.1016/j.physleta.2020.126311}{mSPIRE}].

\bibitem{37} R. Demkowicz-Dobrza\'{n}ski, W. G\'{o}ecki and M. Gu\c{t}\u{a},
Multi-parameter estimation beyond quantum Fisher information
,\href{https://doi.org/10.1088/1751-8121/ab8ef3}{J. Phys. A \textbf{53 }(2020) 363001}
[\href{https://doi.org/10.1088/1751-8121/ab8ef3}{arXiv:2001.11742}]
[\href{https://doi.org/10.1088/1751-8121/ab8ef3}{inSPIRE}].

\bibitem{38} H. Chen, L. Wang and H. Yuan, Simultaneous measurement of
multiple incompatible observables and tradeoff in multiparameter quantum
estimation
,\href{ https://doi.org/10.1038/s41534-024-00894-x}{ npj Quantum Inf. \textbf{10 }(2024)\textbf{\ }98}[\href{ https://doi.org/10.1038/s41534-024-00894-x}{arXiv:2310.11925}][\href{ https://doi.org/10.1038/s41534-024-00894-x}{inSPIRE}].

\bibitem{39} H. Chen, Y. Chen and H. Yuan, Incompatibility measures in
multiparameter quantum estimation under hierarchical quantum measurements,\href{https://doi.org/10.1103/PhysRevA.105.062442}{Phys. Rev. A 105, 062442 (2022)}  [\href{https://doi.org/10.1103/PhysRevA.105.062442}{arXiv:2109.05807v3}][\href{https://doi.org/10.1103/PhysRevA.105.062442}{inSPIRE}].

\bibitem{40} Helstrom CW (1976) Quantum detection and estimation theory.
Academic Press, New York.[\href{https://api.semanticscholar.org/CorpusID:12758217}{inSPIRE}].

\bibitem{41} Hayashi M (2017) Quantum information theory. Springer, Berlin,
Heidelberg.[\href{https://link.springer.com/series/8431}{inSPIRE}].

\bibitem{42} H. P. Yuen and M. Lax, Multiple-parameter quantum estimation
and measurement of nonselfadjoint observables,\href{https://ieeexplore.ieee.org/document/1055103}{IEEE Trans. Inf. Theory 
\textbf{19 }(1973) 740.}[\href{https://ieeexplore.ieee.org/document/1055103}{inSPIRE}].

\bibitem{43} J. Suzuki, Information geometrical characterization of quantum
statistical models in quantum estimation theory,\href{ https://doi.org/10.3390/e21070703}{ Entropy \textbf{21 }(2019)
703.}
[\href{ https://doi.org/10.3390/e21070703}{	arXiv:1807.06990}]
[\href{ https://doi.org/10.3390/e21070703}{inSPIRE}].

\bibitem{44} S. Ragy, M. Jarzyna and R. Demkowicz-Dobrza\'{n}ski,
Compatibility in multiparameter quantum metrology,\href{https://doi.org/10.1103/PhysRevA.94.052108}{Phys. Rev. A \textbf{94 }%
(2016) 052108 } [\href{https://doi.org/10.1103/PhysRevA.94.052108}{arXiv:1608.02634}][\href{https://doi.org/10.1103/PhysRevA.94.052108}{inSPIRE}].

\bibitem{45} Y. Yang, G. Chiribella and M. Hayashi, Attaining the ultimate
precision limit in quantum state estimation, \href{https://doi.org/10.1007/s00220-019-03433Focus to learn more}{ CommunMath. Phys. \textbf{368 }(2019) 223}
[\href{https://doi.org/10.1007/s00220-019-03433-4Focus to learn more}{arXiv:1802.07587}]
[\href{https://doi.org/10.1007/s00220-019-03433-4Focus to learn more}{inSPIRE}].

\bibitem{46} K. Matsumoto, A new approach to the Cram\'{e}r-Rao-type bound
of the pure-state model,\href{https://doi.org/10.1088/0305-4470/35/13/307}{J. Phys. Math. Gen. \textbf{35 }(2002) 3111} [\href{https://doi.org/10.1088/0305-4470/35/13/307}{arXiv:quant-ph/9711008}][\href{https://doi.org/10.1088/0305-4470/35/13/307}{inSPIRE}].

\bibitem{47} M. G. Genoni, M. G. A. Paris, G. Adesso, H. Nha, P. L. Knight
and M. S. Kim, Optimal estimation of joint parameters in phase space,\href{https://doi.org/10.1103/PhysRevA.87.012107Focus to learn more}{Phys.
Rev. A \textbf{87 }(2013) 012107 }[\href{https://doi.org/10.1103/PhysRevA.87.012107Focus to learn more}{arXiv:1206.4867}] [\href{https://doi.org/10.1103/PhysRevA.87.012107Focus to learn more}{inSPIRE}].

\bibitem{48} M. Bradshaw, P. K. Lam and S. M. Assad, Ultimate precision of
joint quadrature parameter estimation with a Gaussian probe, \href{https://doi.org/10.1103/PhysRevA.97.012106}{Phys. Rev. A 
\textbf{97 }(2018) 012106}[\href{https://doi.org/10.1103/PhysRevA.97.012106}{arXiv:1710.04817}] [\href{https://doi.org/10.1103/PhysRevA.97.012106}{inSPIRE}].

\bibitem{49} K. Park, C. Oh, R. Filip and P. Marek, Optimal estimation of
conjugate shifts in position and momentum by classically correlated probes
and measurements
 ,\href{https://doi.org/10.1103/PhysRevApplied.18.014060}{ Phys. Rev. Applied \textbf{18 }(2022) 014060} [\href{https://doi.org/10.1103/PhysRevApplied.18.014060}{arXiv:2203.03348}][\href{https://doi.org/10.1103/PhysRevApplied.18.014060}{inSPIRE}].

\bibitem{50} F. Hanamura, W. Asavanant, S. Kikura, M. Mishima, S. Miki, H.
Terai, M. Yabuno, F. China, K. Fukui, M. Endo and A. Furusawa, Single-shot
single-mode optical two-parameter displacement estimation beyond classical
Limit,\href{https://doi.org/10.1103/PhysRevLett.131.230801}{ Phys. Rev. Lett. \textbf{131 }(2023) 230801}[\href{https://doi.org/10.1103/PhysRevLett.131.230801}{arXiv:2308.15024}] [\href{https://doi.org/10.1103/PhysRevLett.131.230801}{inSPIRE}].

\bibitem{51} J. W. Gardner, T. Gefen, S. A. Haine, J. J. Hope and Y. Chen,
Achieving the fundamental quantum limit of linear waveform estimation, \href{https://doi.org/10.1103/PhysRevLett.132.130801} {Phys.
Rev. Lett. \textbf{132 }(2024) 130801  }[\href{https://doi.org/10.1103/PhysRevLett.132.130801}{arXiv:2308.06253}][\href{https://doi.org/10.1103/PhysRevLett.132.130801}{inSPIRE}].

\bibitem{52} F. Albarelli, J. F. Friel and A. Datta, Evaluating the Holevo
Cram\'{e}r-Rao bound for multiparameter quantum metrology,\href{https://doi.org/10.1103/PhysRevLett.123.200503}  {Phys. Rev. Lett. 
\textbf{123 }(2019) 200503 }[\href{https://doi.org/10.1103/PhysRevLett.123.200503}{arXiv:1906.05724v1}] [\href{https://doi.org/10.1103/PhysRevLett.123.200503}{inSPIRE}].

\bibitem{53} S. K. Chang, M. G. Genoni and F. Albarelli, Multiparameter
quantum estimation with Gaussian states: efficiently evaluating Holevo, RLD
and SLD Cram\'{e}r-Rao bounds ,\href{https://doi.org/10.48550/arXiv.2504.17873}{Phys. A: Math. Theor. 53 385301}[\href{https://doi.org/10.48550/arXiv.2504.17873}{arXiv:2504.17873}]
[\href{https://doi.org/10.48550/arXiv.2504.17873}{inSPIRE}].

\bibitem{54} L. O. Conlon, J. Suzuki, P. K. Lam and S. M. Assad, Efficient
computation of the Nagaoka--Hayashi bound for multiparameter estimation with
separable measurements, \href{https://doi.org/10.1038/s41534-021-00414-1}{npj Quantum Inf. \textbf{7 }(2021) 110}[\href{https://doi.org/10.1038/s41534-021-00414-1}{arXiv:2008.02612}][\href{https://doi.org/10.1038/s41534-021-00414-1}{inSPIRE}].

\bibitem{55} Nagaoka H (2005) A new approach to Cram\'{e}r-Rao bounds for
quantum state estimation, in asymptotic theory of quantum statistical
inference. World Scientific, Singapore.[\href{https://ui.adsabs.harvard.edu/abs/2005atqs.book..100N/abstract}{inSPIRE}].

\bibitem{56} M. Hayashi and Y. Ouyang, Tight Cram\'{e}r-Rao type bounds for
multiparameter quantum metrology through conic programming, 
\href{https://doi.org/10.22331/q-2023-08-29-1094}{Quantum \textbf{%
7 }(2023) 1094 }[\href{https://doi.org/10.22331/q-2023-08-29-1094}{arXiv:2209.05218}] [\href{https://doi.org/10.22331/q-2023-08-29-1094}{inSPIRE}].

\bibitem{57} S. K. Yung, L. O. Conlon, J. Zhao, P. K. Lam and S. M. Assad,
Comparison of estimation limits for quantum two-parameter estimation, 
\href{https://doi.org/10.1103/PhysRevResearch.6.033315}{Phys.Rev. Res. \textbf{6 }(2024) 033315}  [\href{https://doi.org/10.1103/PhysRevResearch.6.033315}{arXiv:2407.12466}] [\href{https://doi.org/10.1103/PhysRevResearch.6.033315}{inSPIRE}].

\bibitem{58} B. Li, L. O. Conlon, P. K. Lam and S. M. Assad, Optimal
single-qubit tomography: Realization of locally optimal measurements on a
quantum computer, \href{https://doi.org/10.1103/PhysRevA.108.032605}{Phys. Rev. A \textbf{108 }(2023) 032605 }  [\href{https://doi.org/10.1103/PhysRevA.108.032605}{	arXiv:2302.05140}]
[\href{https://doi.org/10.1103/PhysRevA.108.032605}{inSPIRE}].

\bibitem{59} M. Zhang, H. M. Yu, H. D. Yuan, X. G. Wang, R. Demkowicz-Dobrza%
\'{n}ski and J. Liu, QuanEstimation: an open-source toolkit for quantum
parameter estimation,\href{https://doi.org/10.1103/PhysRevA.108.032605}{ Phys. Rev. Res. \textbf{4 }(2022) 043057}
[\href{https://doi.org/10.1103/PhysRevA.108.032605}{arXiv:2205.15588}]
[\href{https://doi.org/10.1103/PhysRevA.108.032605}{inSPIRE}].

\bibitem{60} M. Hayashi and Y. Ouyang, Finding the optimal probe state for
multiparameter quantum metrology using conic programming,
\href{https://doi.org/10.1038/s41534-024-00905-x}{npj Quantum Inf. \textbf{10 }(2024)\textbf{\ }111 }[\href{https://doi.org/10.1038/s41534-024-00905-x}{arXiv:2401.05886}]  [\href{https://doi.org/10.1038/s41534-024-00905-x}{inSPIRE}].

\bibitem{61} A. Carollo, B. Spagnolo and D. Valenti, Uhlmann curvature in
dissipative phase transitions,  \href{https://doi.org/10.1038/s41598-018-27362-9}{Sci. Rep. \textbf{8 }(2018) 9852}
[\href{https://doi.org/10.1038/s41598-018-27362-9}{arXiv:1710.07560}][\href{https://doi.org/10.1038/s41598-018-27362-9}{inSPIRE}].

\bibitem{62} Nielsen MA and Chuang IL (2010) Quantum computation and quantum
information. Cambridge University Press, Cambridge. 
[\href{https://doi.org/10.1017/CBO9780511976667}{inSPIRE}].

\bibitem{63} Cram\'{e}r H (1946) Mathematical methods of statistics.
Princeton University Press, Princeton. [\href{https://doi.org/10.1515/9781400883868}{inSPIRE}].

\bibitem{64} C. W. Helstrom, Minimum mean-squared error of estimates in
quantum statistics, \href{https://doi.org/10.1016/0375-9601(67)90366-0}{Phys. Lett. A \textbf{25 }(1967) 101. } [\href{https://doi.org/10.1016/0375-9601(67)90366-0}{inSPIRE}].

\bibitem{65} J. Liu, H. Yuan, X. M. Lu and X. Wang, Quantum Fisher
information matrix and multiparameter estimation,
\href{https://doi.org/10.1088/1751-8121/ab5d4d}{J. Phys. A \textbf{53 }%
(2020) 023001  }[\href{https://doi.org/10.1088/1751-8121/ab5d4d}{arXiv:1907.08037}][\href{https://doi.org/10.1088/1751-8121/ab5d4d}{inSPIRE}].

\bibitem{66} A. S. Holevo, Statistical decision theory for quantum systems,\href{https://doi.org/10.1016/0047-259X(73)90028-6}
{J. Multivar. Anal. \textbf{3 }(1973) 337. }[\href{https://doi.org/10.1016/0047-259X(73)90028-6}{inSPIRE}].

\bibitem{67} Nagaoka H (2005) A generalization of the simultaneous
diagonalization of Hermitian matrices and its relation to quantum estimation
theory. \href{https://doi.org/10.1142/9789812563071_0012}{World Scientific, Singapore. }[\href{https://doi.org/10.1142/9789812563071_0012}{inSPIRE}]

\bibitem{68} R. D. Gill and S. Massar, State estimation for large ensembles,\href{https://doi.org/10.1103/PhysRevA.61.042312}{Phys. Rev. A \textbf{61 }(2000) 042312  }
[\href{https://doi.org/10.1103/PhysRevA.61.042312}{arXiv:quant-ph/9902063}][\href{ https://doi.org/10.1103/PhysRevA.61.042312}{inSPIRE}].

\bibitem{69} M. E. O. Bezerral, F. Albarelli and R. Demkowicz-Dobrzanski,
Simultaneous optical phase and loss estimation revisited: measurement and
probe incompatibility,\href{https://doi.org/10.1088/1751-8121/ade516}{ J. Phys. A: Math. Theor. \textbf{58 }(2025) 265303}
 [\href{https://doi.org/10.1088/1751-8121/ade516}{arXiv:2504.02893}][\href{https://doi.org/10.1088/1751-8121/ade516}{inSPIRE}].

\bibitem{70} L. O. Conlon, J. Suzuki, P. K. Lam and S. M. Assad, The gap
persistence theorem for quantum multiparameter estimation,
 [\href{https://doi.org/10.48550/arXiv.2208.07386}{arXiv:2208.07386v3}][\href{https://doi.org/10.48550/arXiv.2208.07386}{inSPIRE}]

\bibitem{71} A. Das, L. O. Conlon, J. Suzuki, S. K. Yung, P. K. Lam and S.
M. Assad, Holevo Cram\'{e}r-Rao bound: How close can we get without
entangling measurements?, \href{https://doi.org/10.48550/arXiv.2405.09622}{Quantum \textbf{9 }(2025) 1867} [\href{https://doi.org/10.48550/arXiv.2405.09622}{arXiv:2405.09622}][\href{https://doi.org/10.48550/arXiv.2405.09622}{inSPIRE}].

\bibitem{72} L. O. Conlon, J. Suzuki, P. K. Lam and S. M. Assad, Role of the
extended Hilbert space in the attainability of the quantum Cram\'{e}r--Rao
bound for multiparameter estimation,\href{https://doi.org/10.1103/PhysRevLett.91.070402}{Phys. Lett. A \textbf{542 }(2025)
130445 }[\href{https://doi.org/10.1103/PhysRevLett.91.070402}{arXiv:2404.01520}] [\href{https://doi.org/10.1103/PhysRevLett.91.070402}{inSPIRE}].

\bibitem{73} G. Lindblad, On the generators of quantum dynamical semigroups.\href{https://doi.org/10.1142/S1230161223500014}{Commun. Math. Phys. \textbf{48 }(1976) 119} 
[\href{https://doi.org/10.1142/S1230161223500014}{	arXiv:2202.06812}]
[\href{https://doi.org/10.1142/S1230161223500014}{inSPIRE}].

\bibitem{74} F. Benatti, R. Floreanini and M. Piani, Environment induced
entanglement in Markovian dissipative dynamics,\href{https://doi.org/10.1142/S1230161223500014}{Phys. Rev. Lett. \textbf{91 }%
(2003) 070402} [\href{https://doi.org/10.1142/S1230161223500014}{arXiv:quant-ph/0307052}] [\href{https://doi.org/10.1142/S1230161223500014}{inSPIRE}].

\bibitem{75} Z. Zhao and B. Yang, Geometric phases acquired for a two-level
atom coupled to fluctuating vacuum scalar fields due to linear acceleration
and circular motion, \href{https://doi.org/10.1103/PhysRevD.106.036013}{Phys. Rev. D \textbf{106 }(2022) 036013} [\href{https://doi.org/10.1103/PhysRevD.106.036013}{arXiv:2202.10888}]
[\href{https://doi.org/10.1103/PhysRevD.106.036013}{inSPIRE}].

\bibitem{76} J. Suzuki, Explicit formula for the Holevo bound for
twoparameter qubit-state estimation problem,\href{https://doi.org/10.1063/1.4945086}{ J. Math. Phys. \textbf{57 }%
(2016) 042201}[\href{https://doi.org/10.1063/1.4945086}{	arXiv:1505.06437}][\href{https://doi.org/10.1063/1.4945086}{inSPIRE}].

\end{thebibliography}
\end{document}